\documentclass{aa} 
\usepackage{graphicx}
\usepackage{txfonts}
\usepackage{upgreek}
\usepackage{hyperref}
\hypersetup{colorlinks = true, citecolor = {blue}, urlcolor= {blue}}
\def\PGPU{$\varphi-$GPU }
\def\PGRAPE{$\varphi-$GRAPE }

\def\gapprox{\;\rlap{\lower 3.0pt                       % approximately smaller
        \hbox{$\sim$}}\raise 2.5pt\hbox{$>$}\;}
\def\lapprox{\;\rlap{\lower 3.1pt                       % approximately smaller
        \hbox{$\sim$}}\raise 2.7pt\hbox{$<$}\;}

\newcommand{\be}{ \begin{equation} }
\newcommand{\ee}{\end{equation}}

\newcommand{\ben}{\begin{enumerate}}
\newcommand{\een}{\end{enumerate}}

\newcommand{\orcid}[1]{\href{https://orcid.org/#1}{\protect\includegraphics[width=8pt]{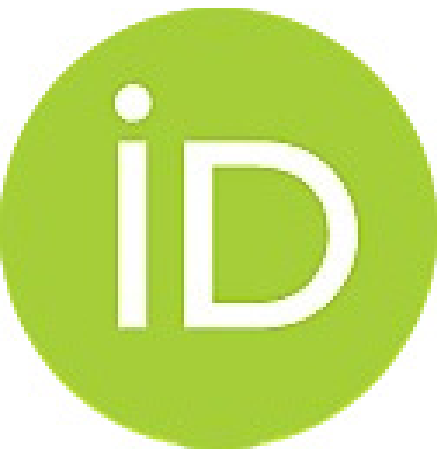}}}
\makeatletter
\renewcommand*\aa@pageof{, page \thepage{} of \pageref*{LastPage}}
\makeatother

\begin{document}
   
\title{Merging of unequal mass binary black holes in non-axisymmetric galactic nuclei}

\author{Peter~Berczik
\inst{1,2,3}\orcid{0000-0003-4176-152X}
\and
Manuel~Arca~Sedda
\inst{4}\orcid{0000-0002-3987-0519}
\and
Margaryta~Sobolenko
\inst{3}\orcid{0000-0003-0553-7301}
\and
Marina~Ishchenko
\inst{3}\orcid{0000-0002-6961-8170}
\and
Olexander Sobodar
\inst{3}\orcid{0000-0001-5788-9996}
\and
Rainer Spurzem
\inst{1,5}\orcid{0000-0003-2264-7203}
}

\institute{Astronomisches Rechen-Institut, Zentrum f\"ur Astronomie, University of Heidelberg, M\"onchhofstrasse 12-14, 69120, Heidelberg, Germany 
           \and
Konkoly Observatory, Research Centre for Astronomy and Earth Sciences, E\"otv\"os Lor\'and Research Network (ELKH), MTA Centre of Excellence, Konkoly Thege Mikl\'os \'ut 15-17, 1121 Budapest, Hungary
           \and
Main Astronomical Observatory, National Academy of Sciences of Ukraine, 27 Akademika Zabolotnoho St., 03143 Kyiv, Ukraine \\
           \email{\href{mailto:berczik@mao.kiev.ua}{berczik@mao.kiev.ua}}
           \and
Physics and Astronomy Department Galileo Galilei, University of Padova, Vicolo dell’Osservatorio 3, I–35122, Padova, Italy
           \and
Kavli Institute for Astronomy and Astrophysics, Peking University, Yiheyuan Lu 5, Haidian Qu, 100871, Beijing, China 
           }
   
\date{Received 27 June 2022 / Accepted 13 July 2022}

\abstract    
{
%\LEt{ Only the first word of titles and subheadings should be capitalized (excluding proper nouns).\ +\ You seem to mostly use UK convention, so I've made the appropriate spelling and punctuation changes.}
In this work, we study the stellar-dynamical hardening of unequal mass supermassive black hole (SMBH) binaries in the central regions of merging galactic nuclei. We present a comprehensive set of direct $N$-body simulations of the problem, varying both the total mass and the mass ratio of the SMBH binary (SMBHB). Simulations 
%\LEt{ A\&A uses the past tense to describe specific methods used in a paper, and the present tense to describe general methods and the findings of recent papers. See Sect. 6 of the language guide https://www.aanda.org/for-authors/language-editing/6-verb-tenses. Please review my edits to ensure this was carried out appropriately throughout your paper.}
were carried out with the \PGPU $N$-body code, which enabled us to fully exploit supercomputers equipped with graphic processing units (GPUs). As a model for the galactic nuclei, we adopted initial axisymmetric, rotating models, aimed at reproducing the properties of a galactic nucleus emerging from a galaxy merger event, containing two SMBHs which were unbound  initially. We found no
%\LEt{ UK convention uses single quotation marks to indicate a special use of a word or phrase.\ +\ I wrote out pc because it isn't directly accompanied by a numeral.} 
'final-parsec problem', as our SMBHs tend to pair and shrink without showing significant signs of stalling. This confirms earlier results and extends them to large particle numbers and 
%\LEt{ The intended meaning isn't clear since "for" is linked to "extends" here.\ If possible, I'd suggest adding a different verb before "for", changing the preposition, or simply deleting "for" depending on the intended meaning.} for 
rotating systems. We find that the SMBHB hardening depends on the binary-reduced mass ratio via a single parameter function. Our results suggest that, at a fixed value for the SMBHB primary mass, the merger time of highly asymmetric binaries is up to four order of magnitudes smaller than the equal-mass binaries. This can significantly affect the population of SMBHs potentially detectable as gravitational wave sources.}
     
\keywords{black holes -- binary black holes --- galactic nuclei -- stellar dynamics}

\titlerunning{Merging of Unequal Mass Binary Black Holes in Non-Axisymmetric Galactic Nuclei}
\authorrunning{Berczik et al.}
\maketitle

%%%%%%%%%%%%%%%%%%%%%%%%%%%%%%%%%%%%%%%%%%%%%%%%%%%%%%%%%%%%%%%%%%%%%%%%%%%%%%%%
\section{Introduction} \label{sec:intro}
%%%%%%%%%%%%%%%%%%%%%%%%%%%%%%%%%%%%%%%%%%%%%%%%%%%%%%%%%%%%%%%%%%%%%%%%%%%%%%%%

According to the hierarchical structure formation scenario in the
%\LEt{ Names and the full version of acronyms do not always require each word to be capitalized. Please refer to Sect 2.2 of the language guide for more details.} 
$\Lambda$~cold dark matter ($\Lambda$CDM) paradigm, bright massive galaxies are supposed to form through mergers and the accretion of smaller galaxies. Supermassive black holes (SMBHs) are commonly observed at the centres of most galaxies, and their mass correlates to several properties of their host galaxy, such as bulge mass \citep{Ho2014} or central velocity dispersion \citep[see e.g.][]{Gultekin09}. Therefore, if the majority of galaxies harbour an SMBH in their centres, SMBH binaries (SMBHBs) 
%\LEt{ Since you also use SMBHBs later on in your paper, I've made this change here.\ I've made the necessary changes where it seemed applicable, but please check my work.}
should represent an unavoidable outcome of the hierarchical formation scenario \citep{Begelman1980}. 

Several examples of resolved SMBHBs with a separation of less than 1~kpc are known. The tightest resolved binary is settled in galaxy 0402+379 and was identified via radio observations achieved with the Very Long Baseline Interferometer \citep[VLBI;][]{Rodriguez2006,Bansal2017}. It contains an SMBH pair candidate with a total mass $\sim 10^8{\rm\;M}_\odot$ and a separation of 7.3~pc. The most reliable and intensively observed SMBHB candidate was placed in a nearby ultraluminous infrared galaxy NGC~6240 \citep{Komossa2003}. Two components (South and North) were resolved in X-ray at a separation of 750~pc. High-resolved observations with the 
%\LEt{ Please remember to spell out acronyms upon first appearance in the abstract and then again beginning with the introduction.}
MUSE instrument revealed the possibility of the third smallest component presence at 
%\LEt{ Cardinal directions (north, south, east, west) should not be capitalized unless forming part of a proper noun +\ Please rephrase if appropriate, possibly to "southern nuclei" or "nuclei in the south".}
south nuclei \citep[\citealt{Kollatschny2020}; for the discussion, see][]{Fabbiano2020}. Using the $M_{\rm BH}-\sigma$ relation, BHs with masses of $3.6\pm0.8\times10^{8}\rm\;M_{\odot}$, $7.1\pm0.8\times10^{8}\rm\;M_{\odot}$, and $9.0\pm0.7\times10^{7}\rm\;M_{\odot}$ were obtained for North, South 1, 
%\LEt{ A\&A uses the serial (Oxford) comma between three or more items in a list to avoid confusion. Also use commas after introductory sentences of three or more words. Commas are not necessary between just two parallel items in a sentence.}
and South 2 components, respectively. 
%\LEt{ Please check the intended meaning hasn't changed here.}
The recent optically and spectroscopically resolved binary in galaxy NGC~7727 has a separation of 500~pc and BH masses  of $1.54^{+0.18}_{-0.15}\times10^{8}\rm\;M_{\odot}$ and $6.33^{+3.32}_{-1.40}\times10^{6}\rm\;M_{\odot}$ \citep{Voggel2022}. The OJ287 is an SMBHB candidate detected via quasi-periodic outbursts for which it has been possible to model the binary separation ($\sim0.05$~pc), eccentricity ($\sim0.6$), and spin \citep[modelled taking even the spin-orbit interaction into account, see][]{Valt2007, Valt2010a, Valt2010b}.

The relativistic in-spiral and final coalescence are driven by gravitational wave (GW) emission, making SMBHBs among the strongest sources to be measured with future GW space satellite missions, such as
%\LEt{ A\&A discourages the use of "like" since it can lead to ambiguity.\ "such as" and "similar to" are preferred alternatives depending on the context.} 
the Laser Interferometer Space Antenna \citep[LISA;][]{Gong2011, Amaro-Seoane2013, Amaro-Seoane2017}, or the TianQin \citep{Luo2016}, Taiji \citep{Ruan2020}, or $\upmu$Ares \citep{Sesana2019} space detectors. It is therefore of paramount importance to understand the astrophysical processes that drive SMBHBs from the unbound, pre-merger state, into the relativistic regime and their associated efficiency. This information can be used to place robust, and physically motivated, constraints on the merger rate of these GW sources and provide a test bed for GW signal templates to be compared with future observations.

However, the large number of existing works focussed on full numerical simulations of 
%\LEt{ Pluralizing modifiers should be avoided, please write either "SMBH pairing and evolution", "of the pairing and evolution of SMBHs", or "SMBHs' pairing and evolution".}
SMBHs pairing and evolution have not fully covered all the important parameters' space yet. In particular, within the hierarchical galaxy formation scenario, the range of mass ratios in galaxy mergers can hugely vary from 1:1 (major mergers) to 1:3, down to minor mergers, that is with mass ratios even below 1:10 \citep[cf. e.g.][]{NJB2006, NJO2009}. Given the correlations between %\LEt{ Idem.\ as note 12, please make the appropriate change.}
galaxies and SMBHs masses, it is thus expected that also SMBHBs should form with a wide range of mass ratios. However, the majority of numerical studies based on $N$-body models have mostly focussed on nearly equal mass binaries. Moreover, the galactic nucleus produced in a galaxy merger event is expected to preserve a non-negligible rotation due to the global angular momentum conservation. The collective motion of stars affected by the rotation of the galactic nucleus can have a non-trivial effect on the SMBHB evolution. In this paper, we expand the simulations presented in \citet{BMSB2006} and \citet{WBSK2014} by exploring a wider range of mass ratios, $0.01-1$, and by also taking the rotating nuclei into account.

Furthermore, the new simulations are based on our improved \PGPU $N$-body code \citep{Berczik2013, SBS2021}, which implements general relativity effects in the form of post-Newtonian corrections to the equations of motion \citep{Sobolenko2017, Khan2016}. This will enable us to follow the SMBH binary evolution as a result of the simultaneous action of stellar hardening and GW emission all the way to final coalescence and allow us to model the GW signal during the final inspiral \citep{KHBJ2013, Sobolenko2017, Khan2018}.

The paper is structured as follows.
%\LEt{ Single-sentence paragraphs aren't allowed, thus the suggested edit.}
In Section~\ref{sec:units} we describe the initial conditions adopted in our $N$-body simulations; in Section~\ref{sec:results} we describe how the modelled SMBHB hardening rate varies across the parameter space explored; and in Section~\ref{sec:summary} we summarize and discuss the main conclusions of our paper. 

%Furthermore, in Appendix~\ref{sec:numerical} we describe the %numerical method.

%%%%%%%%%%%%%%%%%%%%%%%%%%%%%%%%%%%%%%%%%%%%%%%%%%%%%%%%%%%%%%%%%%%%%%%%%%%%%%%%
\section{Units and initial conditions} \label{sec:units}
%%%%%%%%%%%%%%%%%%%%%%%%%%%%%%%%%%%%%%%%%%%%%%%%%%%%%%%%%%%%%%%%%%%%%%%%%%%%%%%%
We analysed a suite of 68 $N$-body simulations comprised of $N$~=~25k, 50k, 100k, 200k, 400k, 1M particles, modelling the evolution of an SMBHB embedded in a rotating galactic nucleus, so as to reproduce the typical environment of the inner regions of a post-merged galaxy. Details about these simulations are provided in \cite{BMSB2006}, \cite{BPBMS2009}, and \cite{WBSK2014}. In this work, we focus on the shrinking of the SMBHB as it interacts with surrounding stars. Because the small $N$ runs (namely 25k and 50k) have a larger uncertainty in the hardening results, we exclude them from the discussion of the further results. We can briefly summarize the main features of our models as follows.

%-------------------------------------------------------------------------%
\begin{figure}[htbp]
  \begin{center}
  \includegraphics[width=\columnwidth]{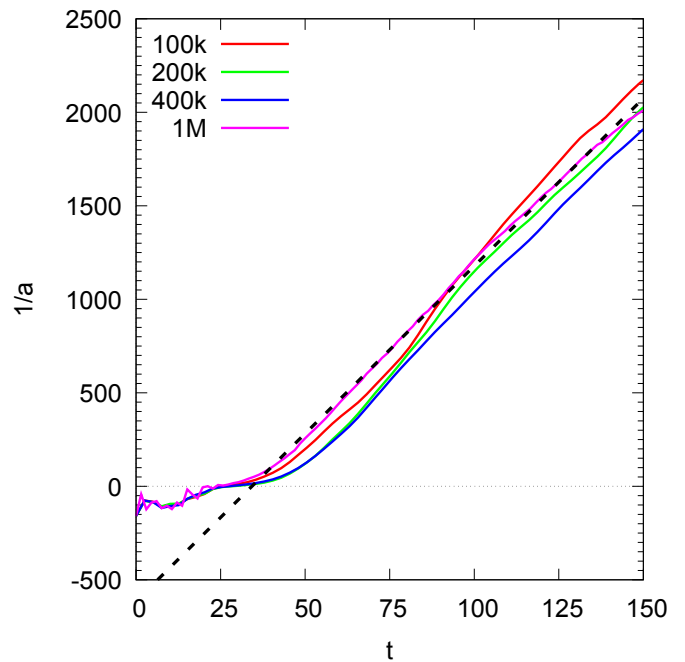}
  \end{center}
  \caption{Inverse of the binary's semi-major axis
        $a$ as a function of time for the four different particles' numbers
        $N$~=~100k,~200k,~400k,~1M model with an SMBH mass ratio of $m_1$:$m_2$=0.01:0.002.
        The black dashed line shows the common corresponding linear fit between
        the time interval $t=50-150$.}
  \label{fig_over_a}
\end{figure}
%-------------------------------------------------------------------------%

The simulations presented in this work have been carried out using the direct $N$-body \PGPU code. Our $N$-body package uses a high order Hermite integration scheme and individual block time steps (the code supports time integration of particle orbits with 
%\LEt{ Write out numerals when lower than eleven and not directly used as a measurement with the unit following. For more details, see Sect 2.7 of the language guide https://www.aanda.org/for-authors/language-editing/2-main-guidelines.}
fourth, sixth, and even eighth order schemes). We refer the more interested readers to a general discussion about different $N$-body codes and their implementation in \cite{BNZ2011, BSW2013}. The \PGPU code is fully parallelized using the MPI library. This code has been newly rewritten 
%from 
%\LEt{ "scratch" in this context is informal, please rephrase.}scratch 
in {\bf \tt C++} and based on the earlier CPU serial $N$-body code \citep[YEBISU;][]{NM2008}. The MPI parallelization was done in the same `$j$'~particle parallelization mode as in the earlier \PGRAPE code \citep{HGM2007}. 

The current version of the \PGPU\footnote{\url{https://github.com/berczik/phi-GPU-mole}}
code uses native GPU support and direct code access to the GPU using the NVIDIA native CUDA library. The code is designed to use different softening parameters for the gravity calculation (if it is required) for different astrophysical components in our simulations, such as SMBHs, dark matter, or stars particles. More details about the public version of the \PGPU code and its performance are presented in \cite{BNZ2011, BSW2013}. The present code has been thoroughly tested and has already been used to obtain important results in our earlier large-scale (up to a few million body) simulations \citep{Khan2018, Zhong2014, LLB2012, Just2012}.

%-------------------------------------------------------------------------%
\begin{figure}[htbp]
  \begin{center}
  \includegraphics[angle=0,width=\columnwidth]{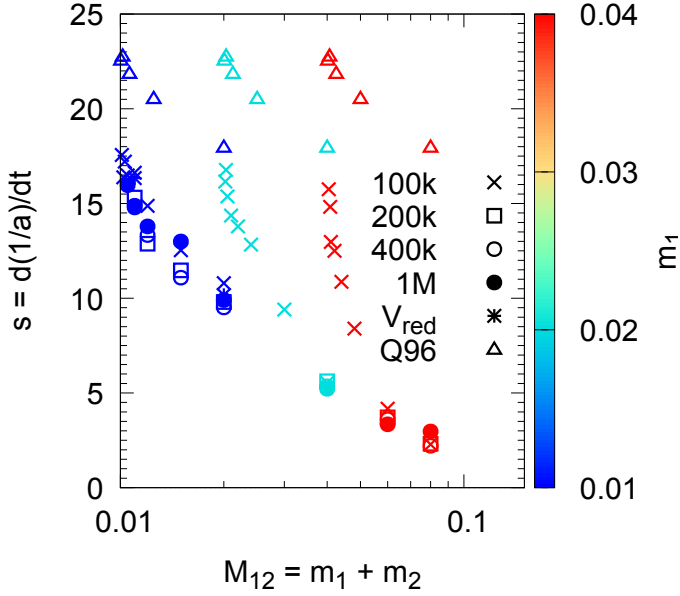}
  \end{center}
  \caption{Hardening rate for SMBH binaries with
  different mass ratios $q = m_2/m_1$ (it is important to 
	%\LEt{ A\&A discourages authors from directly addressing the reader - for example "note that" can be either deleted completely or replaced with "we note that".}
	note $m_1 \ge m_2$) as a function of the binaries' total mass $M_{12} = m_1 + m_2$. Different symbols represent models with different particle numbers $N$ (see Table~\ref{table-mod}), respectively. The open triangle is a parameter $H_{0}$ \citep[see Table~1 in ][]{Quinlan96}. The colours are used to indicate the primary SMBH mass $m_1$.}
  \label{fig_s_vs_M}
\end{figure}
%-------------------------------------------------------------------------%

%-------------------------------------------------------------------------%
\begin{figure}[htbp]
  \begin{center}
  \includegraphics[width=\columnwidth]{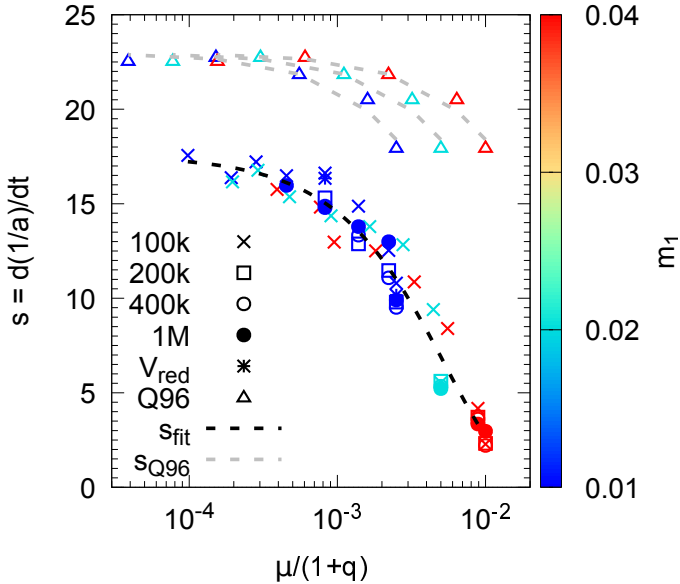}
  \end{center}
  \caption{Hardening rate as a function of the ratio between the reduced mass and the mass ratio, $\mu/(1+q)$. Colour coding marks the mass of the primary component, whereas different symbols refer to different $N$ models. The open triangle is a parameter $H_{0}$ \citep[see Table~1 in ][]{Quinlan96}. The hardening rate was fitted by equation~\ref{eqHar}.}
  \label{rel}
\end{figure}
%-------------------------------------------------------------------------%

All simulations are initially scaled by applying the standard strategy for $N$-body normalization\footnote{\url{https://en.wikipedia.org/wiki/N-body\_units}} \citep{AHW1974}, according to which both the gravitational constant and the total mass of the
stellar systems are set to unity, $G = M = 1$, and the total energy of the system is set to $E=-1/4$.
The galactic nucleus  initially follows a rotating King distribution function \citep[see, e.g.][and references therein]{ES1999}. We set the dimensionless potential well (King parameter) and rotation parameters to $W_0 = 6$ and $\omega_{0}=1.8$, respectively, in all
of our models. At the centre of the nucleus, we placed two SMBHs initially unbound with
a primary mass  $m_1 = (1-4)\times 10^{-2}$ times the nucleus mass and the mass ratio $q = 10^{-4}-1$.
The corresponding range of reduced mass adopted is thus $\mu \equiv m_1 m_2/(m_1+m_2) = 9.9 \times 10^{-3} - 2$.
The total angular momentum vector of both the stellar nucleus and the SMBHs are aligned with the $z$ axis of our coordinate frame. We placed the two SMBHs in the $z=0$ mid-plane with initial coordinate components $x_{1,2}=0$ and $y_{1,2}=\pm 0.3$, where the subscripts denote the two SMBHs. It is important to note that the coordinates of SMBHs are given in $N$-body units and the distance from the centre of the SMBHs' particles roughly corresponds to the influence radius of models in which the SMBHB has a total mass of $2 \times 10^{-2}$. In all the tables and figures, we use these normalized $N$-body units (if not specified directly). 

The initial velocities for the two SMBHs are oriented along the $x$ axis and set to $v_{x;1,2}=\pm V_{\rm circ}$, where $V_{\rm circ}$ is the circular velocity within the stellar background model of nuclei. We note that our choice of initial parameters and scaling implies a circular velocity $V_{\rm circ} \approx 0.7$ in $N$-body units at the initial SMBH positions.

We complement our database of simulations with an extra set of four further models 
%\LEt{ The intended meaning of "in which we assume for the SMBH the initial velocity" is not clear, please rephrase.}
in which we assume for the SMBH the initial velocity $v_{x;1,2}=V_{\rm red}=\pm 0.1 V_{\rm circ} \approx 0.07$. This choice implies that the SMBHs  initially move on more eccentric orbits compared to the other simulations, thus allowing them to pair in a shorter timescale. All the main features of our models are summarized in Table~\ref{table-mod}.

%-------------------------------------------------------------------------%
\begin{table*}[htbp]
%\begin{center}
\centering
\caption{Set of parameters of our model runs. \label{table-mod}}
\begin{tabular}{ccccclllllc}
\hline
\hline
$m_1$ & $m_2$ & $M_{12}$ & $q$ & $\mu$ & 025k & 050k & 100k & 200k & 400k & 1M \\
$10^{-2}$ & $10^{-2}$ & $10^{-2}$ & & $10^{-2}$ & & & & & & \\
(1) & (2) & (3) & (4) & (5) & (6) & (7) & (8) & (9) & (10) & (11) \\
\hline
1 & 0.01 & 1.01 & 0.010 & 0.0099 & ---  & ---  & 450          & ---  & ---  & --- \\
1 & 0.02 & 1.02 & 0.020 & 0.0196 & ---  & ---  & 450          & ---  & ---  & --- \\
1 & 0.03 & 1.03 & 0.030 & 0.0291 & ---  & ---  & 350          & ---  & ---  & --- \\
1 & 0.05 & 1.05 & 0.050 & 0.0476 & ---  & ---  & 350          & ---  & ---  & --- \\
1 & 0.10 & 1.10 & 0.100 & 0.0909 & 250  & 250  & 250$^{\ast}$ & 250  & 250  & 150 \\
1 & 0.20 & 1.20 & 0.200 & 0.1666 & 250  & 250  & 250$^{\ast}$ & 250  & 250  & 150 \\
1 & 0.50 & 1.50 & 0.500 & 0.3333 & 250  & 250  & 250$^{\ast}$ & 250  & 250  & 150 \\
1 & 1.00 & 2.00 & 1.000 & 0.5000 & 250  & 250  & 250$^{\ast}$ & 250  & 250  & 150 \\
\hline             
2 & 0.02 & 2.02 & 0.010 & 0.0198 & ---  & ---  & 450          & ---  & ---  & --- \\
2 & 0.03 & 2.03 & 0.015 & 0.0295 & ---  & ---  & 450          & ---  & ---  & --- \\
2 & 0.05 & 2.05 & 0.025 & 0.0488 & ---  & ---  & 350          & ---  & ---  & --- \\
2 & 0.10 & 2.10 & 0.050 & 0.0952 & ---  & ---  & 350          & ---  & ---  & --- \\
2 & 0.20 & 2.20 & 0.100 & 0.1818 & ---  & ---  & 250          & ---  & ---  & --- \\
2 & 0.40 & 2.40 & 0.200 & 0.3333 & ---  & ---  & 250          & ---  & ---  & --- \\
2 & 1.00 & 3.00 & 0.500 & 0.6666 & 250  & 250  & 250          & 250  & 250  & 150 \\
2 & 2.00 & 4.00 & 1.000 & 1.0000 & 250  & 250  & 250          & 250  & 250  & 150 \\
\hline             
4 & 0.04 & 4.04 & 0.010 & 0.0396 & ---  & ---  & 450          & ---  & ---  & --- \\
4 & 0.08 & 4.08 & 0.020 & 0.0784 & ---  & ---  & 450          & ---  & ---  & --- \\
4 & 0.10 & 4.10 & 0.025 & 0.0976 & ---  & ---  & 350          & ---  & ---  & --- \\
4 & 0.20 & 4.20 & 0.050 & 0.1905 & ---  & ---  & 350          & ---  & ---  & --- \\
4 & 0.40 & 4.40 & 0.100 & 0.3636 & ---  & ---  & 250          & ---  & ---  & --- \\
4 & 0.80 & 4.80 & 0.200 & 0.6666 & ---  & ---  & 250          & ---  & ---  & --- \\
4 & 2.00 & 6.00 & 0.500 & 1.3333 & 250  & 250  & 250          & 250  & 250  & 150 \\
4 & 4.00 & 8.00 & 1.000 & 2.0000 & 250  & 250  & 250          & 250  & 250  & 150 \\
\hline
\end{tabular}
%\end{center}
\tablefoot{Final integration time in $N$-body units. 
Col. 1 -- 2: SMBH masses (primary and secondary, respectively) 
in a $10^{-2}$ model units.
Col. 3: Total mass $M_{12}=m_{1}+m_{2}$ in $10^{-2}$ model units.
Col. 4: Mass ratio $q=m_{2}/m_{1}$, where $m_{2}\geq m_{1}$.
Col. 5: Reduced mass $ \mu \equiv m_1 m_2/(m_1+m_2)$
in $10^{-2}$ model units.
Col. 6 -- 11: Particle number in the stellar galactic nucleus.
$^{\ast}$: in these simulations, we performed two independent sets of runs. After the first set of runs, where the initial orbital velocity of the SMBH was exactly circular, $V_{\rm circ}$, we ran a second set of runs where the initial velocity of the SMBHs was $V_{\rm red}=0.1~V_{\rm circ}$.}
\end{table*}
%-------------------------------------------------------------------------%

%%%%%%%%%%%%%%%%%%%%%%%%%%%%%%%%%%%%%%%%%%%%%%%%%%%%%%%%%%%%%%%%%%%%%%%%%%%%%%%%
\section{Results and discussion} \label{sec:results}
%%%%%%%%%%%%%%%%%%%%%%%%%%%%%%%%%%%%%%%%%%%%%%%%%%%%%%%%%%%%%%%%%%%%%%%%%%%%%%%%
Due to the initial rotation, the stellar distribution undergoes a strong initial bar instability and later forms a rotating triaxial nucleus, as has been discussed in our previous works \citep{BMSB2006, BPBMS2009, WBSK2014}.
The two SMBHs migrate inwards due to the dynamical friction until the point at which they become gravitationally bound.

After the SMBHB formation, continuous interactions with the surrounding field stars cause the shrinking of the orbit, that is to say there is a progressive decrease in the binary semi-major axis $a$. The semi-major axis evolution can be divided into two phases: one during which the binary orbit evolution is driven by both dynamical friction and stellar hardening and, another, dominated by stellar encounters during which the binary shrinking proceeds at a nearly constant rate. This process in basic steps was described and studied in %the 
previous classical works \citep{MV1992, MF2004}. The work of the highly eccentric SMBHB dynamical evolution in the environment of `field stars' %(pro-grade and retro-grade cases) 
presented and described in %the paper
\cite{Iwasawa2011} was especially important; it was shown that the eccentricity increase in SMBH binaries is driven by a combination of a secular effect impinged by the overall nucleus tidal field and a cumulative effect coming from the star-SMBHB scattering in a prograde configuration. In our simulations, these two effects are likely mixed, thus making it hard to discern between them. We postpone any analysis on the binary eccentricity evolution to a forthcoming paper, and we point the interested reader to our previous works for Newtonian simulations \citep{BPBMS2009}.

%%%%%%%%%%%%%%%%%%%%%%%%%%%%%%%%%%%%%%%%%%%%%%%%%%%%%%%%%%%%%%%%%%%%%%%%%%%%%%%%
\subsection{Hardening rate} \label{subsec:hard}
%%%%%%%%%%%%%%%%%%%%%%%%%%%%%%%%%%%%%%%%%%%%%%%%%%%%%%%%%%%%%%%%%%%%%%%%%%%%%%%%

Figure~\ref{fig_over_a} shows the time evolution of the inverse of the binary semi-major axis ($1/a$) in the case of a set of models characterized by $m_1$:$m_2$=0.01:0.002, assuming $N$ = 100k, 200k, 400k, and 1M.
The plot outlines that the hardening follows a nearly linear trend, regardless of the number of particles. Therefore the slope of the $1/a$ curve provides us with an estimate of the binary hardening rate, $s(t) = d(1/a)/dt$. The black dashed line in Figure~\ref{fig_over_a} represents the best-fitting linear relation calculated for the time interval $t = 50 - 150$ ($N$-body units). Coupling the hardening rate with the extra hardening triggered by GW emission enables us to infer the binary merger time \citep[e.g.][]{Gualandris2012,Arca-Sedda2018,G2022}.

Figure~\ref{fig_s_vs_M} shows the dimensionless hardening rate for SMBHBs with different mass ratios $q \equiv m_2/m_1$ as a function of the binary total mass $M_{12} = m_1 + m_2$ for models with different particle numbers $N$ (see Table~\ref{table-mod}) and different values for the mass of the primary SMBH. The plot outlines how, at a fixed  SMBHB total mass, a heavier primary -- that is to say a lower mass ratio -- determines larger hardening. Similarly, at a fixed primary mass, we see that heavier binaries -- that is to say heavier companions -- are associated with a smaller hardening rate. This implies that the efficiency of stellar hardening is maximized for low-mass-ratio SMBHBs with a heavy primary mass. This result holds regardless of the number of particles used, thus confirming that is a physical effect rather than a numerical one. 

Taking advantage of scattering experiments, \cite{Rasskazov2017} and \cite{Rasskazov2019} recently showed that for low mass ratios ($q < 10^{-2}$), the hardening rate is proportional to $(1+q)^2/q$. We find a similar dependence in our full $N$-body models, namely that the hardening is tightly linked to a simple combination of the binary primary mass and mass ratio defined as follows:
\begin{equation}
f(q,\mu) \equiv \frac{m_1 q}{(1+q)^2} = \frac{m_2}{(1+q)^2} = \frac{\mu}{(1+q)}.
\end{equation}

Figure~\ref{rel} shows the relation between the hardening rate and the $\mu/(1+q)$ quantity. We find that this relation is well fitted by the following simple exponential formula:
\begin{equation}
%s = {\rm A} \exp(- {\rm B} \mu/(1+q)),
s_{\rm fit} = \frac{d}{dt}\Bigl(\frac{1}{a}\Bigr) = {\rm A} \exp\Bigl(-\frac{{\rm B}\mu}{1+q}\Bigr),
\label{eqHar}
\end{equation}
with A=$17.5\pm0.3$ and B=$186.7\pm8.8$. We note that this relation implies that the secondary mass and the total mass ratio play the most effective role in shaping the SMBHB evolution. Also we fitted the data from Table~1 \citep{Quinlan96} for systems with masses of $m_{1}$=0.01, 0.02, and 0.04 and obtained B$_{0.01}$=88.6, B$_{0.02}$=44.3, and  B$_{0.04}$=22.1  accordingly when  A=23.0 for all models.

The fitting formula provided above enables us to link the binary hardening rate to the masses of the binary components. Figure~\ref{rel} compares our fitting formula with the simulated data for the 52 largest $N$ models considered here. We see that in passing from $\mu/(1+q)=10^{-4}$ to $10^{-2}$, the hardening rate decreases by a factor of 5. At a fixed primary mass, this implies that highly asymmetric binaries ($q\ll1$) tend to be characterized by a more effective hardening compared to the case of nearly equal mass SMBHBs. This might crucially affect the merger efficiency of SMBHBs in minor galaxy mergers, which are expected to be the main contributors to the population of SMBHBs with a mass ratio from 1:10-1:3, especially at high redshift \citep[e.g.][]{Callegari2011}. Our results might have interesting consequences for GW astronomy since primary SMBHs with a mass from $10^6-10^7\rm\; M_{\odot}$ merging with a small companion might be bright GW sources for LISA \citep[][]{Amaro-Seoane2017} and similar space-based detectors such as TianQin \citep{Luo2016}, Taiji \citep{Ruan2020}, or $\upmu$Ares \citep{Sesana2019}.

%%%%%%%%%%%%%%%%%%%%%%%%%%%%%%%%%%%%%%%%%%%%%%%%%%%%%%%%%%%%%%%%%%%%%%%%%%%%%%%%
\subsection{Merger efficiency} \label{subsec:merg}
%%%%%%%%%%%%%%%%%%%%%%%%%%%%%%%%%%%%%%%%%%%%%%%%%%%%%%%%%%%%%%%%%%%%%%%%%%%%%%%%
We can use these results to place constraints on the SMBH merger efficiency. The long-term evolution of an SMBH binary can be written as follows \citep[e.g.][]{Gualandris2012,Arca-Sedda2018,G2022}:
\begin{eqnarray}
\frac{da}{dt} &=& \frac{da}{dt}\Big|_{\ast} + \frac{da}{dt}\Big|_{\rm GW},\\
\frac{de}{dt} &=& \frac{de}{dt}\Big|_{\ast} + \frac{de}{dt}\Big|_{\rm GW},
\label{eqGua}
\end{eqnarray}
where the asterisk identifies stellar hardening, whereas GW refers to the hardening driven by GW emission.

While the stellar hardening term in our simulations is a direct consequence of continuous star-SMBHB interactions and can be described through the adimensional hardening rate described in the previous section, the GW term for the evolution of semi-major axes is given by the following \citep{Peters1963,Peters1964a,Peters1964b}:
\begin{eqnarray}
\label{EqSe2}
\frac{da}{dt}\Big|_{\rm GW} &=& -\frac{64\beta}{5}\frac{F(e)}{a^3},\\
\beta &=& \frac{G^3}{c^5 m_1m_2(m_1+m_2)}, \\
F(e)  &=& (1 - e^2)^{-7/2}\Bigl(1+\frac{73}{24}e^2 + \frac{37}{96}e^4\Bigr).
\end{eqnarray}
The eccentricity evolution regulated by stellar encounters and GW emission, respectively, is given by the following:
\begin{eqnarray}
\label{eqEa}
\frac{de}{dt}\Big|_{\ast} &=& \frac{K}{a}\frac{da}{dt}, \\
\label{eqE}
\frac{de}{dt}\Big|_{\rm GW} &=& -\frac{304\beta}{15}\frac{e D(e)}{a^4},\\
\label{eqDe}
D(e) &=& (1-e^2)^{5/2}\left(1+\frac{121}{304}e^2\right), \\
\label{eqK}
K &=& \frac{de}{d\ln(1/a)},
\end{eqnarray}
where $K$ is the eccentricity growth rate \citep{Quinlan96}. As shown in several recent works, the $K$ parameter depends, non-trivially, on the MBH masses and the environment \cite[see e.g.][]{WBSK2014,bonetti20}. Due to this, in the following we neglect equation~\ref{eqEa}, thus implicitly assuming that the star-SMBHB interaction does not alter the binary eccentricity significantly \cite[see e.g. Fig. 3 in][]{WBSK2014}.

We solved the system of differential equations~\ref{EqSe2}~--~\ref{eqK}  numerically.\ We also adopted the scaling relation obtained for the adimensional hardening rate derived in equation~\ref{eqHar}, integrating the evolution of SMBH binaries down to the merger.

To obtain clues about the role played by our stellar hardening recipe on the formation of SMBH mergers, we created a sample of 2,000 SMBH binary mergers as follows. Firstly, we selected the primary and companion SMBH mass between $10^4-10^9\rm\;M_{\odot}$, adopting a uniform distribution in logarithmic values. We assumed that the SMBH host nucleus has a mass extracted from between $10^{1.5}$ and $10^4$ times the SMBH mass \citep[the values are in the range from the Milky Way to the classical bulges;][]{KH2013}, 
also adopting a uniform distribution in logarithmic values in this case. 

%-------------------------------------------------------------------------%
\begin{figure}
  \begin{center}
  \includegraphics[width=0.98\linewidth]{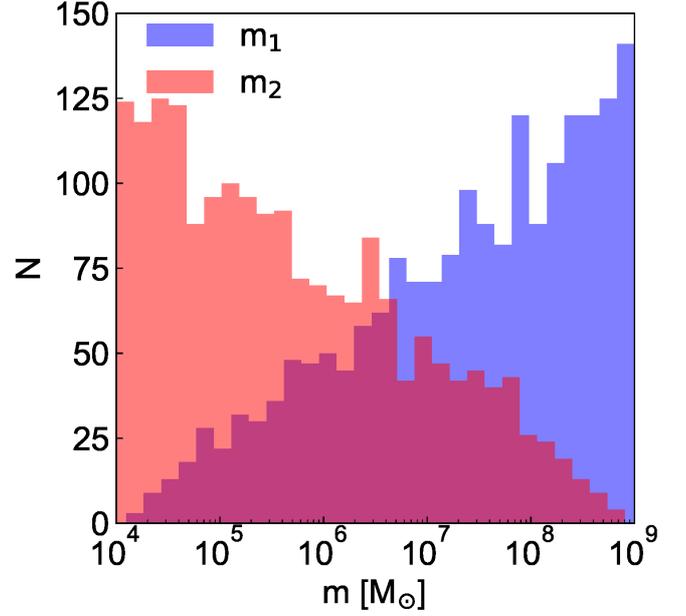}
  \end{center}
  \caption{Distribution of the random realization component masses $m_{1,2}$ ($m_{2}\geq m_{1}$) for merging systems (subsection~\ref{subsec:merg}). The mass range in the middle (cherry colour) shows the overlapping masses.}
  \label{hist}
\end{figure}
%-------------------------------------------------------------------------%

In the Figure~\ref{hist}, we show the random realizations SMBH individual masses $m_{1,2}$. As we can see, our random sampling thoroughly cover the observed range of SMBHs. 
%\LEt{ or "of the SMBH." depending.} 
This implies nuclei with masses in the range $M_{\rm b} = (3.5\times 10^5 - 10^{13})\rm\; M_{\odot}$. For each pair, we calculated the hard-binary semi-major axis $a_{\rm h} = G(m_1+m_2)/2\sigma^2$, evaluating the galaxy velocity dispersion from the $M_{\rm SMBH} -\sigma$ relation \citep{Gultekin09, KH2013}. We started our integration assuming that the binary semi-major axis equals the hard-binary separation and we drew the eccentricity from a thermal distribution\footnote{We also tried to use a uniform distribution and found no significant differences in the results.} \citep{Jeans1919,Ambartsumian1937,Heggie1975,Kroupa2008}.

\begin{figure}
  \begin{center}
  \includegraphics[width=0.98\linewidth]{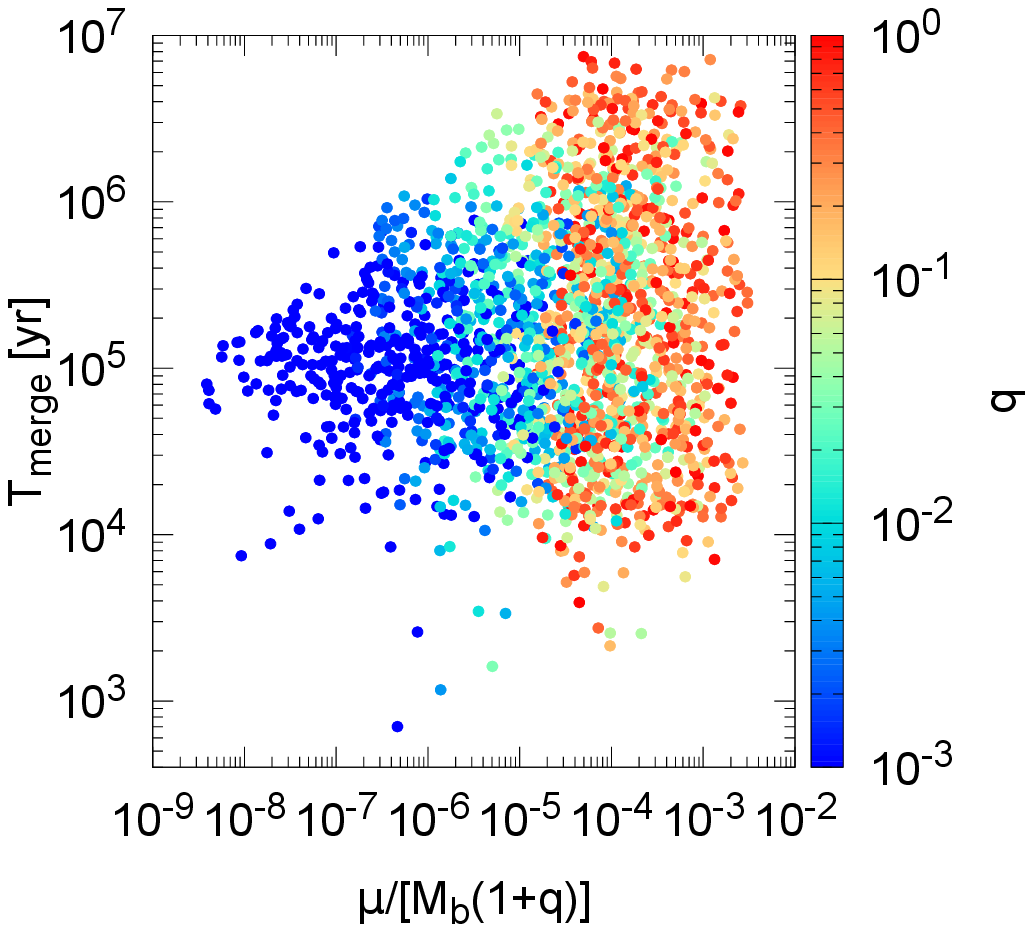}
  \includegraphics[width=0.98\linewidth]{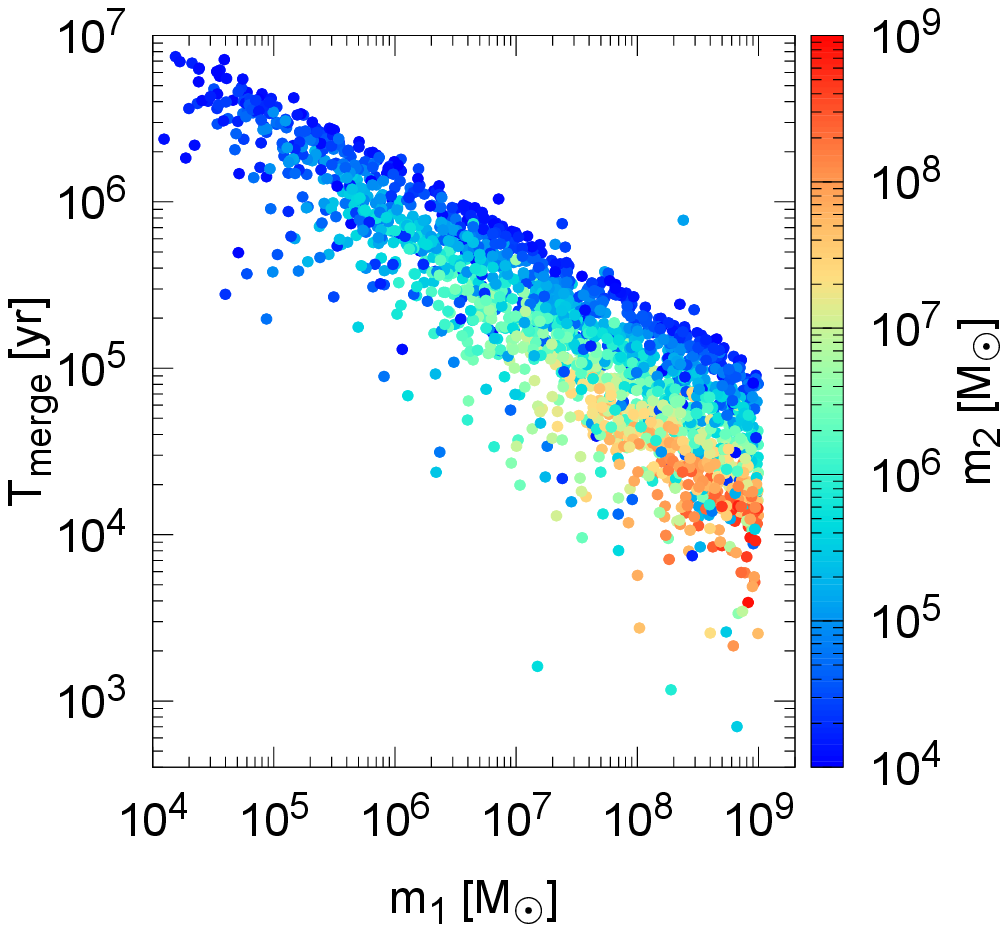}
  \end{center}
  \caption{\textit{Top panel:} 
%	\LEt{ If possible, please add a general title without an initial article before  "Top panel" here and anywhere else where one is not provided for figure or table legends.}
	Merger time $T_{\rm merge}$ as a function of the $\mu/[M_{\rm b}(1+q)]$ parameter conveniently scaled to the galaxy nucleus mass $M_{\rm b}$. The colour coding identifies the binary mass ratio. {\textit{Bottom panel:} Merger time $T_{\rm merge}$ as a function of the primary mass $m_{1}$, where colour coding identifies the secondary mass $m_{2}$.}}
  \label{smbhmer-rot}
\end{figure}

In Figure~\ref{smbhmer-rot}, we show $T_{\rm merge}$ as a function of the $\mu/[M_{\rm b}(1+q)]$ parameter (varying the binary mass ratio and primary mass) and the individual masses of the SMBHB. Looking at the two panels in the figure highlights the role of stellar hardening in the SMBHB evolution, leading the actual merger time to the range of $\sim$10$^7$~yr, which is many orders of magnitude larger as a simple merger timescale estimation based on the Peters-Mathews formalism \citep{Peters1963,Peters1964a,Peters1964b}. At a fixed SMBH primary mass, our results show that the minimization of the merger time is achieved for binaries with lighter SMBH companions, thus with lower mass ratios. This is due to the fact that lower mass ratios imply lower values for the $\mu/[M_{\rm b}(1+q)]$ parameter, which in turn implies a larger hardening in terms of absolute values. The efficient hardening driven by the rotating nucleus thus enables SMBHBs with mass ratios of $q=10^{-4}-1$ and component masses of $m_{1,2} = 10^{4-9}\rm\;M_{\odot}$ (Fig.~\ref{hist}) to merge over timescales, that is $T_{\rm merge} \approx 10^3-10^7$~yr.

%-------------------------------------------------------------------------%
%-------------------------------------------------------------------------%

%%%%%%%%%%%%%%%%%%%%%%%%%%%%%%%%%%%%%%%%%%%%%%%%%%%%%%%%%%%%%%%%%%%%%%%%%%%%%%%%
\section{Summary} \label{sec:summary}
%%%%%%%%%%%%%%%%%%%%%%%%%%%%%%%%%%%%%%%%%%%%%%%%%%%%%%%%%%%%%%%%%%%%%%%%%%%%%%%%
We have presented a systematic study of the evolution of asymmetric SMBH binaries harboured in a non-axisymmetric, dense stellar galactic nucleus. To be able to perform this large systematic set of direct $N$-body simulations, we have developed a new high-performance computing code \PGPU that enables one to fully exploit current and next generations high-performance GPU clusters.

Our simulations suggest that the binary hardening rate for unequal mass SMBHBs does not depend on the number of particles $N$, at least in the regime investigated here. The stellar hardening in our modelled SMBHBs is sufficiently effective to drive the SMBHB
%\LEt{ You may want to write "drive SMBHBs towards" instead here depending.} 
towards coalescence in a short timescale of about a gigayear \citep{BPBMS2009}.

For a fixed mass of the primary SMBH, that is the more massive one, we find a significant increase in the hardening rate for smaller masses of the secondary (i.e. lighter) SMBH. On the other hand, for a fixed mass of the secondary SMBH, we find the hardening rate to be proportional to the primary's mass.

We find a tight relation between the hardening rate and the binary total mass $M_{12}$ and reduced mass $\mu$, which suggests that the SMBHB shrinkage process can be described by a two-parameters' relation. Smaller values of $M_{12}$ and $\mu$ result in an increase in the hardening rate. The hardening rates in our models support predictions of three-body scatter experiments \citep[see e.g.][]{HF1980, Hills1983, Rasskazov2017, Rasskazov2019}, which suggests a scaling between the hardening rate and binary mass ratio.

%%%%%%%%%%%%%%%%%%%%%%%%%%%%%%%%%%%%%%%%%%%%%%%%%%%%%%%%%%%%%%%%%%%%%%%%%%%%%%%%
\begin{acknowledgements}
%%%%%%%%%%%%%%%%%%%%%%%%%%%%%%%%%%%%%%%%%%%%%%%%%%%%%%%%%%%%%%%%%%%%%%%%%%%%%%%%
We thank our colleagues Andreas Just, Christian Boily and Long Wang for their significant help and fruitful discussions during the preparation of the present work.

This work was supported by the Deutsche Forschungsgemeinschaft (DFG, German Research Foundation) Project-ID 138713538, SFB 881 ("The Milky Way System") and by the Volkswagen Foundation under the grant No. 97778. 

We thank the Gauss Centre for Supercomputing e.V. (\url{www.Gauss-centre.eu}) for providing computing time through the John von Neumann Institute for Computing (NIC) on the GCS Supercomputers JUWELS and JUWELS-Booster \citep[JSC;][]{JUWELS2021} in Germany.

PB and MI acknowledge the support from the Ministry of Education and Science of Ukraine under the Ukrainian - French collaborative grant M/2-16.05.2022. 

PB and MI thanks the Partenariats Hubert Curien and Campus France for financial support through the DNIPRO programme no. 46814ZL, and the College de France for the award of a PAUSE short-term 
visiting scholarship tenable at the Observatoire astronomique 
at the Universite de Strasbourg. We also thanks its Director, 
Pierre-Alain Duc, for making our stay in Observatoire astronomique 
de Strasbourg possible, during which this work was being finalized.

PB, MI, MS, and OS acknowledges by support under the special programme of the NRF of Ukraine "Leading and Young Scientists Research Support" - "Astrophysical Relativistic Galactic Objects (ARGO): life cycle of active nucleus",  No. 2020.02/0346.

The work of PB was also supported by the Volkswagen Foundation under the special stipend No. 9B870 (2022).

MS acknowledges the support under the Fellowship of the National Academy of Sciences of Ukraine for young scientists 2020-2022. 

MAS acknowledges financial support from the Alexander von Humboldt Foundation and the Federal Ministry for Education and Research for the research project "The evolution of black holes from stellar to galactic scales", and funding from the European Union’s Horizon 2020 research and innovation programme under the Marie Skłodowska-Curie grant agreement No. 101025436 (project GRACE-BH, PI: Manuel Arca Sedda). 

RS and PB cordially acknowledge hospitality and support by National Astronomical Observatories of Chinese Academy of Sciences in Beijing, China.

%%%%%%%%%%%%%%%%%%%%%%%%%%%%%%%%%%%%%%%%%%%%%%%%%%%%%%%%%%%%%%%%%%%%%%%%%%%%%%%%
\end{acknowledgements}
%%%%%%%%%%%%%%%%%%%%%%%%%%%%%%%%%%%%%%%%%%%%%%%%%%%%%%%%%%%%%%%%%%%%%%%%%%%%%%%%

%%%%%%%%%%%%%%%%%%%%%%%%%%%%%%%%%%%%%%%%%%%%%%%%%%%%%%%%%%%%%%%%%%%%%%%%%%%%%%%%
\bibliographystyle{mnras}
\bibliography{bbh-mu}
%%%%%%%%%%%%%%%%%%%%%%%%%%%%%%%%%%%%%%%%%%%%%%%%%%%%%%%%%%%%%%%%%%%%%%%%%%%%%%%%

%%%%%%%%%%%%%%%%%%%%%%%%%%%%%%%%%%%%%%%%%%%%%%%%%%%%%%%%%%%%%%%%%%%%%
\end{document}